\documentclass[nofootinbib,superscriptaddress,a4paper,twocolumn, floatfix]{revtex4-2}
\usepackage[english]{babel}
\usepackage[utf8]{inputenc}

\usepackage{amsthm}
\usepackage{mathtools}
\usepackage{physics}
\usepackage{xcolor}
\usepackage{graphicx}
\usepackage[left=23mm,right=13mm,top=35mm,columnsep=15pt]{geometry} 
\usepackage{algpseudocode}

\usepackage{adjustbox}
\usepackage{placeins}
\usepackage[T1]{fontenc}
\usepackage{lipsum}
\usepackage{csquotes}
\usepackage{tabularx}

\usepackage[pdftex, pdftitle={Article}, pdfauthor={Author}]{hyperref} % For hyperlinks in the PDF
\usepackage{xcolor}
\hypersetup{
    colorlinks,
    linkcolor={red!50!black},
    citecolor={blue!50!black},
    urlcolor={blue!80!black}
}
\usepackage{booktabs}
\usepackage{multirow}
\usepackage{listings}
\usepackage{xcolor}
\usepackage{caption}

\begin{document}

\title{Comparing Classical and Quantum Ground State Preparation Heuristics}

\author{Katerina Gratsea}
\email[E-mail:]{gratsea.katerina@gmail.com}
\affiliation{ICFO - Institut de Ci\`{e}ncies Fot\`{o}niques, The Barcelona Institute of Science and Technology, Av. Carl Friedrich Gauss 3, 08860 Castelldefels (Barcelona), Spain}
\affiliation{Zapata AI, Inc.}

\author{Jakob S. Kottmann}
\affiliation{Institute for Computer Science, University of Augsburg, Germany}

\author{Peter D. Johnson}
\affiliation{Zapata AI, Inc.}

\author{Alexander A. Kunitsa}
\affiliation{Zapata AI, Inc.}

\date{\today} % Leave empty to omit a date

\begin{abstract}
One promising field of quantum computation is the simulation of quantum systems, and specifically, the task of ground state energy estimation (GSEE).
Ground state preparation (GSP) is a crucial component in GSEE algorithms, and classical methods like Hartree-Fock state preparation are commonly used. However, the efficiency of such classical methods diminishes exponentially with increasing system size in certain cases. In this study, we investigated whether in those cases quantum heuristic GSP methods could improve the overlap values compared to Hartree-Fock. Moreover, we carefully studied the performance gain for GSEE algorithms by exploring the trade-off between the overlap improvement and the associated resource cost in terms of T-gates of the GSP algorithm. Our findings indicate that quantum heuristic GSP can accelerate GSEE tasks, already for computationally affordable strongly-correlated systems of intermediate size. These results suggest that quantum heuristic GSP has the potential to significantly reduce the runtime requirements of GSEE algorithms, thereby enhancing their suitability for implementation on quantum hardware.
\end{abstract}

\maketitle

\section{Introduction}

The simulation of quantum systems is one of the more promising applications of quantum computers~\cite{cao2019quantum}.
In particular, many methods have been proposed to solve the ubiquitous task of ground state energy estimation~\cite{poulin2018quantum, lin2022heisenberg, low_depth_GSEE_Peter}.
Unfortunately, the quantum resources needed to solve this task for industrial applications is many millions of physical qubits
\cite{kim2022fault, goings2022reliably} and the computations can take days to months to run.
The value of large-scale quantum computing will depend on the degree to which these costs can be reduced~\cite{evidence_paper, katabarwa2023early}.

For the task of ground state energy estimation in both classical and quantum computing, a critical subtask is \emph{ground state preparation}.
Examples of approximate ground state preparation include Hartree-Fock  (HF) \cite{echenique2007mathematical}, configuration interaction (CI) \cite{sherrill1999configuration}, and density-matrix renormalization (DMRG) \cite{wouters2014chemps2, zhai2023block2}.
The performance of the ground state energy estimation method depends on the quality of an approximate ground state, known as a trial or ansatz state~\cite{Google_overlaps, Zhang_2022, LD_new, LD_tables};
the greater the magnitude of the inner product or \emph{overlap} between the ansatz and the true ground state, the more efficient the ground state energy estimation can be.

This dependence has led to an increased
interest in approximate ground state preparation (GSP) algorithms~\cite{Google_overlaps, fomichev2023initial}.
The Hartree-Fock state preparation, one of the most established classical methods for state preparation, is often the baseline method for ground state preparation in quantum algorithms for quantum chemistry. 
HF state preparation works well for many systems~\cite{Google_overlaps}, but in strongly correlated cases, the overlap can decrease exponentially with the system size~\cite{evidence_paper}. 
Recent works~\cite{acc_critera, Andrew_2023} discuss the cost-benefit ratio of quantum GSP over HF for the task of GSEE, while other works focus on more expensive and accurate classical GSP methods~\cite{chromium, evidence_paper, jellium}. If quantum GSP methods offer a better trade-off between performance and resource cost over HF, then the same holds for most other classical chemistry methods~\cite{new_Google}.

Regardless of the recent progress in the field, there is still ongoing research on how quantum GSP methods could significantly improve the task of GSEE. A lot of progress has been made in GSP algorithmic development, but it remains unclear how costly these quantum heuristic algorithms are in regard to GSEE algorithms. To tackle such questions, numerical simulations of quantum heuristic GSP are necessary to reach efficient ground state preparation as emphasized in~\cite{evidence_paper}. In this work, we carefully evaluate the trade-off of the resource cost in terms of the gate count for an improved overlap value of the studied quantum heuristic GSP over classical methods~\cite{acc_critera}.

Towards this goal, we perform numerical simulations and resource estimations motivated by the following questions:
\begin{itemize}
    \item To what degree
    can GSP algorithms improve the ground state overlap when the corresponding HF overlap has small values? (see Sec.~\ref{H6} and Fig.~\ref{fig:H6_plot}) 
    \item Could quantum heuristic algorithms continue to improve the ground state overlap beyond that of HF as we increase the system size? (see Sec.~\ref{Hn} and Fig.~\ref{fig:Hn_SPA})
    \item Assuming the quantum GSP algorithms as subroutines of GSEE, to what degree do they improve the performance while maintaining moderate computational cost? (see Sec.~\ref{speed-ups} and Fig.~\ref{fig:Hn_ratios})
\end{itemize}

We focus on two \textit{heuristic} quantum GSP methods, the variational quantum eigensolver (VQE) method known as separable pair approximation (SPA)~\cite{SPA} and the ground state booster~\cite{Wang_2022}. The recent work~\cite{Andrew_2023} explored similar aspects of the aforementioned questions but for the near-optimal ``non-heuristic'' GSP method of Lin and Tong~\cite{Lin_2020}. Given the near-optimal provable performance guarantees of such methods, we choose to investigate heuristic GSP techniques which are not necessarily limited by the same lower bounds and might empirically perform better. A study to perform a direct comparison between ``non-heuristic'' and heuristic GSP methods could be the focus of a future work.

Here, we perform a detailed numerical analysis and resource estimation to compare heuristic ground state preparation techniques.
Quantitatively, we aim to determine the degree to which different ground state preparation methods reduce the runtime of ground state energy estimation.
This involves estimating the overlaps between the ansatz states and the ground state as well as estimating the number of quantum gates used to implement the circuits; as we will explain, both components play a role in determining the runtime reduction.
As is common in fault-tolerant resource estimation \cite{psi_ratios, goings2022reliably, beverland2022assessing, Andrew_2023}, we will assume that the number of non-Clifford operations, specifically T-gates, governs the runtime of the quantum circuit and therefore measure runtimes in terms of circuit T-gate counts.

As a system of study, we use the 1-D hydrogen chains, which are good candidates for strongly correlated multi-electron systems~\cite{SPA, H10}. Such systems capture many central themes of modern condensed-matter physics~\cite{Hn_properties} and the essential features of the many-electron problems in real materials~\cite{Hn_real_life}. Interestingly, even for such simple models as linear hydrogen chains, research is still ongoing; only recently, fundamental ground state properties were computed in Ref.~\cite{Hn_properties}.

This paper is organized as follows. In Sec.~\ref{H6}, we focus on the first question posed on how quantum GSP methods can improve small values of the HF overlap. In Sec.~\ref{Hn}, we explore the performance of quantum GSP methods as a function of the system size. In Sec.~\ref{speed-ups}, we analyze the speed-ups that quantum GSP methods could give for the task of GSEE. Finally, we present the conclusions and outlook in Sec.~\ref{conclusions}.

\section{Quantum GSP performance over HF}\label{H6}

In this section, we focus on the first question posed in the introduction: \textit{To what degree can GSP algorithms improve the ground state overlap when the corresponding HF overlap has small values?} 
Both the SPA and booster algorithms have the potential to boost the overlap value even when the value of the initial HF overlap is very small, though the gate costs of the methods may need to increase as the overlap decreases
\cite{SPA, Wang_2022}.

The SPA algorithms are variations of the VQE algorithm optimized using a separable pair approximation (SPA), which assumes that a (closed shell) $N$-electron system could be described by a wavefunction of $N/2$ electron pairs. 
Each electron pair is represented by a wavefunction restricted to a disjoint subset of orbitals. Once combined with specific circuit compilation strategies, they can give classically tractable circuit classes with very short circuit depths. Specifically, the circuit depths are constant with the system size $N$ and scale linearly with the basis size~\cite{SPA}. Moreover, as with other variational methods like adiabatic state preparation, the SPA does not necessarily 
require
high overlap between its input state (e.g., Hartree-Fock) and the true ground state, as it can, in principle, prepare states orthogonal to Hartree-Fock due to its unitary nature. 

The booster algorithm implements a function $f$ (i.e., a Gaussian) of a Hamiltonian $H$, which suppresses the high-energy eigenstates of the Hamiltonian and enhances the low-energy ones in the expansion of an initial state. These functions $f$ are referred to as boosters~\cite{Wang_2022}. Even though some initial non-zero overlap is necessary for the booster algorithm to perform, it was demonstrated that it can convert an increase in circuit depth 
into
an increase in overlap value, which in general, is not a feature of VQE-type algorithms. The aforementioned question becomes quite interesting for the booster algorithm if we limit the depth proxy $D$ (or truncation level, see App.~\ref{App:booster} for more details) of the truncated Fourier expansion $f_D \left( H \right)$ of the booster function $f$ to a certain value. Then, boosting the overlap becomes more challenging but also practical for implementations on early fault-tolerant quantum hardware \cite{katabarwa2023early}. 

\begin{figure}[ht]
    \centering
    \includegraphics[width=1.0\linewidth]{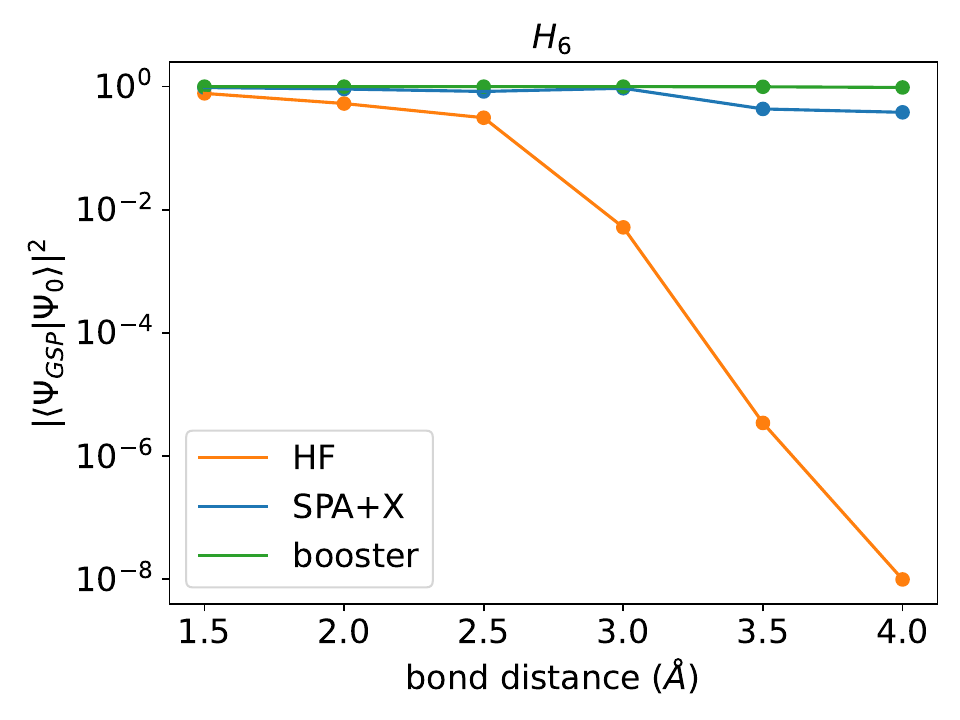}
    \caption{Here we plot the HF, SPA+X with approximates orbital rotations on different levels: $X\in \text{GS, GAS}$, and booster $|\langle \Psi_{GSP} | \Psi_{0} \rangle|^2$ for the $H_6$ system with MRA-PNO basis set for an increasing bond distance.}
    \label{fig:H6_plot}
\end{figure}

In Fig.~\ref{fig:H6_plot}, we compare the behavior of the SPA algorithm to restricted HF for the system of $H_6$ with $8$ spin-orbitals or qubits in an adapted basis (MRA-PNO basis)~\cite{Kottmann_2021} for a range of bond distances (i.e. the spacing between hydrogen atoms). As already discussed, the SPA algorithm can improve the performance over the initial HF overlaps even when the HF fidelity has a very small value, i.e., $10^{-7}$. Interestingly though, while the HF overlaps decrease exponentially as we increase the bond distance, the overlaps obtained with the SPA remain approximately constant. Even better, in the case of the booster algorithm, the squared overlaps are approximately equal to one for all bond distances. 

For each bond distance shown in Fig.~\ref{fig:H6_plot}, we used different circuit variants of the SPA algorithm~\cite{SPA} and plotted the variants that give the maximum overlap. Specifically, for the $H_6$ molecule, we have the following variants SPA+GAS, SPA, SPA, SPA+GS, SPA+GS, and SPA+GS (referred to as "SPA+X" in Fig.~\ref{fig:H6_plot}) for the corresponding bond distances $1.5, 2.0, 2.5, 3.0, 3.5, 4.0 \textup{~\AA}$. The three variants supplement the SPA with unitary singles excitations: S adds singles from occupied to virtual orbitals, GS adds single excitations between all orbitals, and GAS approximates the singles by neglecting Pauli-$Z$ operations in the Jordan-Wigner encoding.

\section{Performance of quantum GSP algorithms versus system size}~\label{Hn}

Next, we investigate the second question posed in the introduction: \textit{Could quantum heuristic algorithms improve the ansatz overlap in terms of the exponential decrease with the system size?}

To this end, we explore the behavior of the hydrogen chains $H_n$ with $R(H-H)=3.0 \textup{~\AA}$ in terms of the overlaps as a function of the system size $n \in \{2,4,6,8\}$. We use the MRA-PNO basis for $H_n$, which is closer to being complete compared to the minimal basis and is a better proxy for real-world use cases. Therefore, we have $H_n$ for $n \in \{2,4,6,8\}$ with $4, 8, 12, 16$ spin-orbitals or qubits respectively in an adapted basis~\cite{Kottmann_2021}. 

\begin{figure}[ht]
    \centering
    \includegraphics[width=1.0\linewidth]{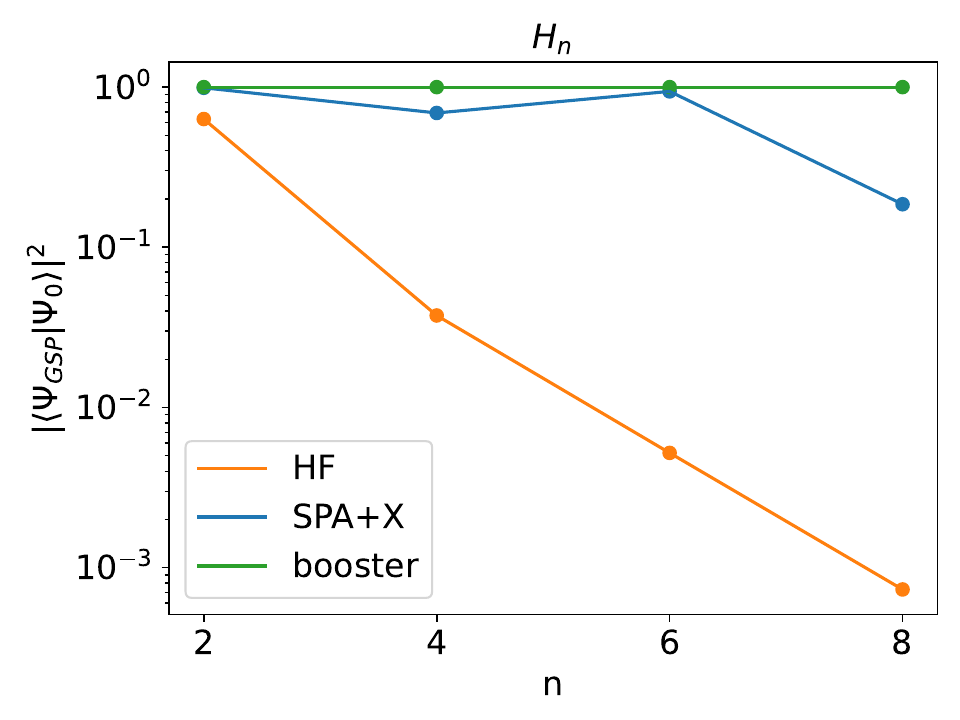}
    \caption{Here we plot the HF, SPA+X, and booster overlap squared for the $H_n$ systems with MRA-PNO basis set and bond distance $R(H-H)=3.0 \textup{~\AA}$ for increasing system size $n$.}
    \label{fig:Hn_SPA}
\end{figure}

In Fig.~\ref{fig:Hn_SPA}, we plot the HF, SPA-variants, and booster fidelities for the hydrogen chains $H_n$. In App.~\ref{app:sto-3g}, we plot the aforementioned fidelities for the same systems but with STO-3G basis set for comparison following~\cite{kottmann2022molecular} (see Fig.~\ref{fig:Hn_sto-3g_SPA}). In Fig.~\ref{fig:Hn_SPA}, we have used the following SPA-variants (referred to as "SPA+X"): SPA, SPA+S, SPA+GAS, SPA+GS for $n$ equal to $2, 4, 6, 8$, respectively. 

While the HF fidelities drop exponentially with the system size $n$, the SPA algorithm gives more consistent performance. Interestingly though, the booster algorithm gives the maximum fidelity for all system sizes $n$. Potentially, for the case of $n=8$, the SPA algorithm performance could be further improved. 

Generally, the SPA algorithm optimizes the orbitals to give the best energy in the chosen electron pair model which was approximated through pair natural orbitals in the original work~\cite{SPA}. Some molecular instances though, require different shapes of orbitals to be optimal for SPA wavefunctions and the success of the orbital optimization depends on good initial guesses, which is, for example, relevant for $H_n=8$ in Fig.~\ref{fig:Hn_SPA}.

Moreover, the SPA-variant could be further optimized since the SPA+GS depends on the initial guess of ansatz parameters, taken to be "random-close-to-zero" throughout this work, but with suitable intuition as in~\cite{kottmann2022molecular}, the fidelity can be increased to around 90\%. Extensive numerical analysis, to this end, goes beyond the scope of this work, which aims to analyze state-of-the-art procedures that can be applied in a semi-automatic way.

Another important factor for VQE algorithms relates to the number of individual measurements $M$ that affect the runtime linearly and grow inversely proportional to the square of the total energy estimation error. This error can be expressed as $\eta+\epsilon$, where $\epsilon$ refers to the ansatz expressivity error, while $\eta$ to the sampling error with $\eta \leq \epsilon$~\cite{Gonthier_2022}. In the worst case, $ \eta = \epsilon $ which suggests that $M$ grows as $1/\left(2\epsilon \right)^2)$.

Generally getting the energy error within $\epsilon$ does not guarantee that the fidelity error will be within the desirable precision~\cite{mayer2003simple, saad2011numerical}. But in certain cases, like the studied $H_6$ system we end up with an acceptable error in both the energy ($\epsilon=17.9mHa$) and fidelity ($F=0.94$) estimations. Specifically, the energy estimation of the VQE algorithm within $\epsilon=17.9mHa$ is acceptable in scenarios where QPE will be applied and improve the energy estimation to within chemical accuracy. Next, the efficiency criteria introduced in~\cite{acc_critera} are satisfied which renders the VQE algorithm studied here as an acceptable quantum GSP over HF. This is because the fidelity provides a high success rate and a significant speedup in the total runtime of the GSEE algorithm.

Usually, chemical accuracy $\epsilon_0 = 1.6mH$ is the target for the energy estimation error $\epsilon$ of a VQE algorithm, but as discussed in the example above a larger energy error, i.e. $\epsilon=17.9mHa$, might be acceptable. In that case, we 
are afforded
a significant reduction in the number of measurements $M_0/M = \epsilon^2/ \epsilon_0^2 = 100$ 
required 
for a single energy evaluation of the VQE algorithm. The decreased sampling rates might negatively affect the gradient computations of VQE~\cite{Sweke_2020, Gonthier_2022}, but in the studied VQE algorithm the SPA angles could be optimized classically and avoid the aforementioned issue.

\section{Speed-ups of quantum GSP algorithms over HF}\label{speed-ups}

In this section, we focus on the last question posed in the introduction: \textit{Assuming the quantum GSP algorithms as subroutines of GSEE, what speed-ups could they offer?} To this end, we first apply the acceptability criteria introduced in the recent work of~\cite{acc_critera}. According to this work, to determine whether to accept or reject a GSP over the HF we need to carefully examine the cost-benefit ratio, i.e., the resource cost needed to gain an increased overlap value.

In Table~\ref{table:RE}, we present the circuit depths of the studied SPA-variant circuits after compiling the unitaries containing single- and double-excitation gates to single-qubit Pauli rotations $R$ and CNOT gates (see App.~\ref{app:comp_details}). Then, each Pauli rotation is assumed to be implemented as a Cliffort+T circuit following the suggestion in Ref.~\cite{relation} giving the total T-gate count of  
\begin{equation}\label{eqTgates}
    \text{T-count} \approx 3\times R \times \log_2{\left(\dfrac{1}{\delta}\right)},
\end{equation}
where $\delta$ is the necessary precision per gate for operating $R$ gates. As described in App. \ref{App:failure_tolerance}, we choose $\delta$ through the failure tolerance $\delta_C$ of the GSP circuit.
that we are willing to tolerate. If we set the circuit failure tolerance to $\delta_C= 0.001$ for all $n$, 
we get the respective values of $\delta$ (see Table~\ref{table:RE} and App.~\ref{App:failure_tolerance} for more details). Note that the main conclusions will not significantly depend on the value of the failure tolerance.

In Table~\ref{table:RE}, we also give the T-gate counts of the method among the SPA variants with the fewest T-gates per circuit.
Moreover, we report the T-gate counts of the SPA+GS as a proxy for more advanced methods where explicit representations of the orbital rotations in the circuit are necessary, as 
for example in~\cite{kottmann2022molecular} or in case the following GSEE would require the Hamiltonian to be in a different orbital representation. As shown in table~\ref{table:RE}, the $H_2$ system remains on the SPA level as it is already represented exactly. 

Next, we estimate the T-gate counts $T_{GSEE}$ of the GSEE algorithm for the $n$ system size by using the OpenFermion resource estimation module~\cite{openfermion}. Specifically, we get the estimated resources in terms of Toffoli gate count for single factorized QPE as described in Ref.~\cite{SF_1, SF_2, PRXQuantum.2.030305}. To convert the Toffoli gate count to T-gate we assumed $\text{Toffoli}=4 \times \text{T-gate}$~\cite{Tof_to_Tgates}.

\begin{table}[h]
    \centering
    \caption{The lower ($T_{SPA}$) and upper ($T_{SPA+GS}$) bounds on the T-gate counts of the SPA circuits for $H_n$. Also, we give the T-gate counts of the GSEE algorithm ($T_{GSEE}$).}
    \label{table:RE}
    \begin{tabular}{|c|c|c|c|c|}
        \hline
        $H_n$ & $T_{SPA}$ & $T_{SPA+GS}$  &  $T_{GSEE}$\\
        \hline
        $H_2$ &  $21$ & $21$    &  $9.6 \times 10^5 $ \\
        \hline
        $H_4$ &  $46$   &     $4.2 \times 10^3$    &  $1.3 \times 10^7$ \\
        \hline
        $H_6$ &  $72$  &     $1.1 \times 10^4$    & $5.2\times 10^7$ \\
        \hline
        $H_8$ &  $100$  &     $2.2 \times 10^4$    & $1.6\times 10^8 $\\
        \hline
    \end{tabular}
\end{table}

\begin{figure}[ht]
    \centering
    \includegraphics[width=1.0\linewidth]{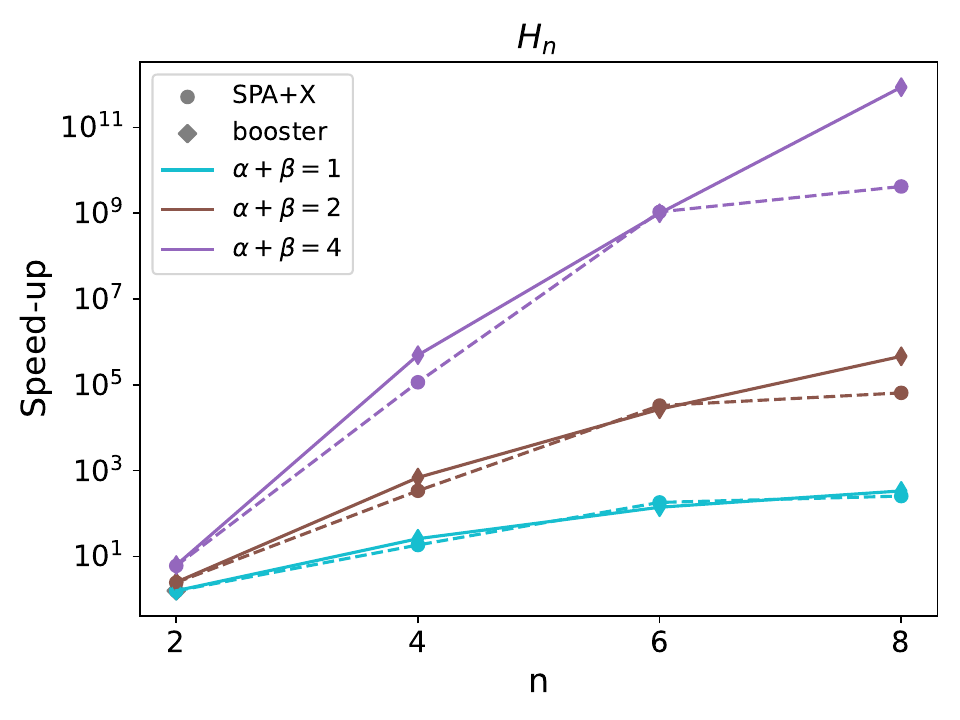}
    \caption{Here we plot the speed-up ratios $T_0/T$ for SPA+X and booster algorithms over HF of $H_n$ system for an increasing system size $n$ and bond distance $R(H-H)=3.0\textup{~\AA}$. The $\alpha$ and $\beta$ parameters are scaling parameters for GSEE algorithms}
    \label{fig:Hn_ratios}
\end{figure}

Comparing the T-gate counts of the SPA and GSEE algorithms shown in Table~\ref{table:RE}, we observe that $ T_{GSEE} + T_{SPA+X} \approx T_{GSEE}$. Then, as discussed in~\cite{acc_critera}, the acceptability criterion becomes $ 1 <\left(\frac{\gamma}{\gamma_0}\right)^{\alpha+\beta} $, where $\gamma$ and $\gamma_0$ correspond to SPA and HF overlap, respectively. Here $\gamma^{\beta}$ 
is the scaling of the gate complexity of the ground state energy estimation circuit (excluding the state preparation)
while $\gamma^{\alpha}$ 
is the scaling of the number of circuit repetitions
needed to ensure the success for the given GSEE algorithm~\cite{LD_tables}. Given the criterion's dependence on the sum of these parameters and taking a few example algorithms from~\cite{LD_tables}, we consider the GSEE algorithms to be represented by
$\alpha + \beta = \{1,2,4\}$. For example, $\alpha+\beta=1,2 \text{ and } 4$ correspond to the works of~\cite{ab1, ab2} and~\cite{ab4}, respectively. Generally, a lower sum indicates a more-performant GSEE algorithm.
Figure \ref{fig:Hn_ratios} demonstrates that
the SPA-variants are acceptable for all of the aforementioned values of $\alpha+\beta$ over the HF state preparation for the $H_n$ systems shown in Fig.~\ref{fig:Hn_SPA}. 

According to~\cite{acc_critera} the speed-up ratio becomes $\dfrac{T_0}{T} = \left(\dfrac{\gamma}{\gamma_0}\right)^{\alpha+\beta}$, where $T_0$ and $T$ refers to running the GSEE algorithm with HF and SPA-variants as subroutines for ground state preparation, respectively. In Fig.~\ref{fig:Hn_ratios}, we plot the aforementioned speed-up ratios for the values of $ \alpha+\beta \in [1, 2, 4] $. 

Next, we perform the same analysis, but with the low-depth booster as the GSP method. In that case, the circuit depth (measured as the number of accumulated controlled-$\exp\left( 2 \pi i H \right)$  operations) is $2D$ where $D$ is the depth proxy (or truncation level) of the truncated Fourier expansion of $f_D \left( H \right)$ (see App.~\ref{App:booster}). 

Given that the booster operation can be implemented with the linear combination of unitaries (LCU), we use a Trotter decomposition with one Trotter step to estimate the lowest cost in terms of Pauli rotations $R$ and CNOT gates (see Table~\ref{table:comp_details}). We can convert the Pauli rotations to T-gates using Eq.~\ref{eqTgates} (see $T_{K=1}$ in Table~\ref{table:RE_booster}). The cost of the booster circuits in T-gate count is $T_{B}= 2D K T_{K=1}$ (see Table~\ref{table:RE_booster}), where $K$ is the number of total Trotter steps required. A more detailed analysis of the booster resource estimation can be found in App.~\ref{App:booster}.

The recent work of~\cite{Trotter_steps} empirically found that the required number of Trotter steps $K$ for the 1-D Hydrogen chain is approximately equal to $10$ and we discuss in more detail the applicability of this result in our case in App.~\ref{App:Trotter_steps}. In this work, we fix $D=10$, since it is sufficient for the booster to give overlap values equal to $1.0$ with the respective success probability $P_{succ}$ shown in Table~\ref{table:RE_booster}.  The $T_{GSEE}$ is the same as in Table~\ref{table:RE}.

\begin{table}[h]
    \centering
    \caption{We present the important factors for the booster resource estimation analysis: the success probability ($P_{succ}$) and the total T-gate counts ($T_{B}$) along with the T-gate counts $T_{K=1}$ of the Trotter decomposition with one Trotter step $K=1$ and of the GSEE algorithm ($T_{GSEE}$) for $H_n$.}
    \label{table:RE_booster}
    \begin{tabular}{|c|c|c|c|c|}
        \hline
       $H_n$ & $P_{succ}$ & $T_{K=1}$ &  $T_{B}$ & $T_{GSEE}$ \\
        \hline
        $H_2$ & 0.46 & $9.3 \times 10^2$  & $1.9 \times 10^5$ & $9.6 \times 10^5$\\
        \hline
        $H_4$ & 0.027 & $5.6 \times 10^4$ & $1.1 \times 10^7$  & $1.3 \times 10^7$ \\
        \hline
        $H_6$ & $1.1\times 10^{-4}$ & $3.5 \times 10^5$  & $7.0 \times 10^7$ & $5.2 \times 10^7$ \\
        \hline
        $H_8$ & $1.1\times 10^{-4}$ & $1.3 \times 10^6$ & $2.6 \times 10^8$ & $1.6 \times 10^8 $ \\
        \hline
    \end{tabular}
\end{table}

Following the work of~\cite{acc_critera}, we can calculate the speed-up ratio $\dfrac{T_0}{T}$. Here we drop the $\Tilde{O}$ by using the T-gate counts presented in Table~\ref{table:RE_booster} for both the booster and GSEE algorithm. Thus, the acceptability criteria can be written as 
\begin{equation}
    \dfrac{\mathcal{T}_0}{\mathcal{T}}=\frac{\left(\dfrac{T_{GSEE}}{\Tilde{\epsilon} \gamma_0^{\alpha+\beta}}\right)}{\left(\dfrac{1}{\gamma^\alpha}\left(\dfrac{T_{B} }{P_{s u c c}}+\dfrac{T_{GSEE}}{\Tilde{\epsilon} \gamma^\beta}\right)\right)},
\end{equation}
where $T_{B}$ and $T_{GSEE}$ are the T-gate counts of the booster and GSEE algorithm with target accuracy $\Tilde{\epsilon}$. 
In Fig.~\ref{fig:Hn_ratios}, we plot the speed-up ratios for the systems presented in Fig.~\ref{fig:Hn_SPA} with the corresponding T-gate counts presented in Table~\ref{table:RE_booster} (see App.~\ref{App:overlaps} for the respective overlap values). According to the work of~\cite{acc_critera}, since all speed-up ratios are larger than one, the booster is acceptable over HF for all system sizes presented in Fig.~\ref{fig:Hn_ratios}.

\section{Discussion and Outlook}\label{conclusions}

% Findings

We performed simulations on small strongly correlated systems $H_n$ with an increasing bond distance $R(H-H)$ and system size $n$. These systems show a rapid decrease in fidelity between the restricted Hartree-Fock state and the ground state. On the contrary, both quantum GSP methods studied, i.e., SPA and booster, give fidelities close to $1.0$. Similar performance is observed in Fig.~\ref{fig:Hn_SPA}, where we study different hydrogen chains with increasing system size $n\in[2,4,6,8]$. 

Following the work of~\cite{acc_critera}, we also focused on the cost-benefit ratio for the quantum GSP methods for the task of GSEE. In Fig.~\ref{fig:Hn_ratios}, we plot the speed-up ratios gained by using the SPA and booster algorithm over the HF for the studied $H_n$ systems with increasing size $n$. The $\alpha+\beta$ determines the dependence of the GSEE algorithm on the ground state overlap $\gamma$, i.e. $\left( \gamma \right)^{-\left(\alpha + \beta \right)}$~\cite{LD_tables}. Therefore, as suggested by Fig.~\ref{fig:Hn_ratios}, the smaller the value of $\alpha+\beta$, the smaller the speed-up ratio gained by the quantum GSP algorithm over the HF.  

% Context

To put these results into context, recently there have been ongoing research on whether quantum GSP methods are necessary over classical ones. 
Different works have stressed the difficulties with classical GSP methods and the related problem of vanishing overlaps for certain chemical systems~\cite{nobel-lecture, evidence_paper}. But other works have proposed that classical state preparation methods should not be a limiting factor in phase estimation even for large or strongly-correlated systems~\cite{goings2022reliably, Google_overlaps}. 

Our work suggests that quantum heuristic GSP methods could be beneficial over classical ones by reducing the runtime requirements of the GSEE algorithms in which they are used. Even for computationally affordable (i.e. requiring few resources and time to complete) strongly-correlated systems of intermediate size, i.e. $H_n$ with $n \in [4,6,8]$, we observe significant speed-ups over HF method for both SPA and booster algorithm. Already for $n=4$ we report an order of magnitude speed-up for both studied GSP methods, while for most instances of $n$ and $\alpha+\beta$ we have many orders of magnitude runtime improvement. 

Most literature that explores runtime improvements of quantum algorithms on quantum chemistry tasks has focused on large-scale systems and materials~\cite{psi_ratios, evidence_paper, goings2022reliably}. However, recent works on classical chemistry methods establish the linear hydrogen chain as a benchmarking system for numerical simulations~\cite{Hn_properties, H10, Hn_real_life}. Thus, despite its simplicity, the hydrogen chain incorporates a rich set of physical and chemical properties~\cite{Hn_properties} and could be used as a benchmarking set to help develop quantum computing applications for quantum chemistry.

% Strengths and limitations

In principle, the success of the studied quantum GSP algorithms for the hydrogen chains could have been predicted but not the exact values of the scaling.
For example, other systems that could exhibit similar behavior with the 1-D hydrogen chains are linear and cyclic organic molecules with pi-bonds (for example, alkenes or aromatic hydrocarbons $C_n H_n$).

% What's next?
There are several directions following this line of exploration that we leave for future work.
First, it would be important to assess the GSEE runtime savings in cases where the HF state preparation may have an associated cost.
In this paper, we assumed that HF state preparation is a cost-free operation on quantum hardware, which is valid in second quantization of the Hamiltonian but not necessarily valid with other approaches such as first quantization \cite{su2021fault}. 
It would be important for future work to re-evaluate the conclusions of this paper using such methods where the Hartree-Fock preparation incurs a T-gate cost.

Second, the analysis could be extended to hydrogen chains with an increased system size $N$~\cite{fomichev2023initial} and with different bond distances and geometries~\cite{H10}. The studied bond distance $3.0\AA$ for the linear hydrogen chain corresponds to a Mott insulator phase~\cite{Hn_properties, Hn_real_life}, and the correlations can be characterized by a spin-1/2 Heisenberg chain. Decreasing the bond distance below $1.7\AA$~\cite{Hn_properties, Hn_real_life} will allow us to study significant subjects in quantum material physics, such as metal-insulator transitions and magnetism.

% The "so what"? Implicate, speculate, recommend.

As ground-state energy estimation is one of the most promising tasks for realizing on quantum computers, our work sheds light on how quantum heuristic GSP algorithms could reduce the runtime requirements of GSEE, and, therefore, make them more realistic for implementations on quantum hardware. Our work highlights that runtime improvements could be achieved for the task of GSEE by quantum heuristic GSP methods over HF for computationally affordable strongly correlated systems of intermediate size, like linear $H_n$ with $n \in [4,6,8]$, which can capture interesting physical phenomena. This work aims to elucidate the landscape of methods that might someday be used to solve utility-scale problems in quantum chemistry.

\bibliography{main}

\section{Acknowledgement}

We would like to thank Athena Caesura, Jerome Gonthier, Max Radin, Paul Cazeaux, and Agnieszka Miedlar for constructive and inspiring conversations.

This work was performed while K.G. was a research intern at Zapata AI, Inc. Part of this research was performed while K.G. was visiting the Institute for Pure and Applied Mathematics (IPAM), which is supported by the National Science Foundation (Grant No. DMS-1925919). K.G. acknowledges support from the European Union's Horizon 2020 research and innovation programme under the Marie Skłodowska-Curie grant agreement No. 847517. 

ICFO group acknowledges support from ERC AdG NOQIA; Ministerio de Ciencia y Innovation Agencia Estatal de Investigaciones (PGC2018-097027-B-I00/10.13039/501100011033, CEX2019-000910-S/10.13039/501100011033, Plan National FIDEUA PID2019-106901GB-I00, FPI, QUANTERA MAQS PCI2019-111828-2, QUANTERA DYNAMITE PCI2022-132919,  Proyectos de I+D+I “Retos Colaboración” QUSPIN RTC2019-007196-7); MICIIN with funding from the European Union NextGenerationEU (PRTR-C17.I1) and by Generalitat de Catalunya;  Fundació Cellex; Fundació Mir-Puig; Generalitat de Catalunya (European Social Fund FEDER and CERCA program, AGAUR Grant No. 2021 SGR 01452, QuantumCAT \ U16-011424, co-funded by ERDF Operational Program of Catalonia 2014-2020); Barcelona Supercomputing Center MareNostrum (FI-2023-1-0013); EU (PASQuanS2.1, 101113690); EU Horizon 2020 FET-OPEN OPTOlogic (Grant No 899794); EU Horizon Europe Program (Grant Agreement 101080086 — NeQST), National Science Centre, Poland (Symfonia Grant No. 2016/20/W/ST4/00314); ICFO Internal “QuantumGaudi” project; European Union’s Horizon 2020 research and innovation program under the Marie-Skłodowska-Curie grant agreement No 101029393 (STREDCH) and No 847648  (“La Caixa” Junior Leaders fellowships ID100010434: LCF/BQ/PI19/11690013, LCF/BQ/PI20/11760031,  LCF/BQ/PR20/11770012, LCF/BQ/PR21/11840013). Views and opinions expressed are, however, those of the authors only and do not necessarily reflect those of the European Union, European Commission, European Climate, Infrastructure and Environment Executive Agency (CINEA), nor any other granting authority.  Neither the European Union nor any granting authority can be held responsible for them.

\appendix
\section*{Appendix}

\section{STO-3G}\label{app:sto-3g}

\begin{figure}[ht]
    \centering
    \includegraphics[width=1.0\linewidth]{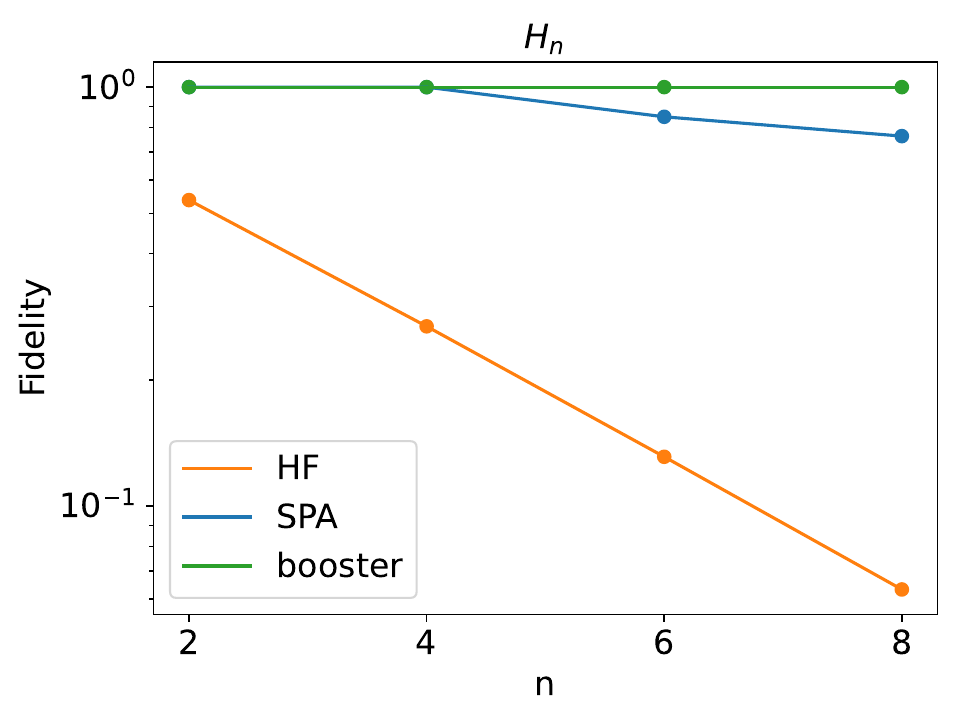}
    \caption{Here we plot the HF, SPA and booster overlap squared for the $H_n$ systems with STO-3G basis set with an increasing system size $n$.}
    \label{fig:Hn_sto-3g_SPA}
\end{figure}

Here, we explore the behaviour of the $H_n$ with the STO-3G basis set in terms of the ansatz overlap as a function of the system size $n \in \{2,4,6,8\}$. In Fig.~\ref{fig:Hn_sto-3g_SPA}, we see that HF fidelities drop faster than SPA, which on the contrary gives reasonable high fidelity values for all studied $n$.

\section{Booster resource estimation analysis}~\label{App:booster}

Following the analysis of the booster algorithm~\cite{Wang_2022}, the booster operation $f(H)$ of the given Hamiltonian $H$ could be implemented by obtaining a Fourier approximation of $f$ and implementing a linear combination of unitaries (LCU) method as
\begin{equation}
    f_D(x)=\int_{-D}^D \hat{f}(\xi) e^{i 2 \pi x \xi} d \xi.
\end{equation}
According to the aforementioned work, the proxy depth of this operation could be estimated as $2D$. Next, the integral could be further discretized as

\begin{equation}
    f_{D, N}(x)=\frac{D}{N} \sum_{j=-N}^{N-1} \hat{f}\left(\xi_j\right) e^{i 2 \pi x \xi_j},
\end{equation}

with $\xi_j=(j+1 / 2) \frac{D}{N} \text { for } j=-N,-N+1, \ldots, N-1$. Next, assuming a Gaussian booster $exp\left( -a x^2 \right)$ the above equation becomes

\begin{equation}\label{function}
    f_{D, N ; a}(x)=\frac{D}{N} \sqrt{\frac{\pi}{a}} \sum_{j=-N}^{N-1} e^{-\frac{\left(\pi \xi_j\right)^2}{a}} e^{2 \pi i x \xi_j},
\end{equation}

where $\xi_j=(j+1 / 2) D / N$, and $N$ is sufficiently large so that $f_a(x) \approx f_{D ; a}(x) \approx f_{D, N ; a}(x)$. Finally, we can apply a Trotter decomposition to the operation $e^{2 \pi i H \xi_j}$ of the given Hamiltonian $H$ for one Trotter step to estimate the Pauli rotations, and in turn, the T-gate count $T_{K=1}$. Usually, a required number of Trotter steps $K$ is necessary to reach the desired precision. Thus, the total number of T-gates for the booster algorithm is given by $T = 2D K T_{K=1}$.

Finally, following the work of~\cite{acc_critera}, the booster algorithm requires a number of repetitions $1/P_{succ}$ to ensure their success, where the success probability $P_{succ}$ is given by 
\begin{equation}
    p_{s u c c}\left(f_{T, N}\right) \approx \frac{\left\langle\psi\left|f^2(H)\right| \psi\right\rangle}{f^2(0)} = \left\langle\psi\left|f^2(H)\right| \psi\right\rangle,
\end{equation}
since $f^2(0)=1$ for the Gaussian booster.

\section{Number of Trotter steps}~\label{App:Trotter_steps} 

The work of~\cite{Trotter_steps} estimates the necessary number of Trotter steps for $1$-D Hydrogen chains to chemical accuracy. To apply this analysis to the operation $e^{2 \pi i H \xi_j}$ discussed in the previous section and Eq.~\eqref{function}, we need that the maximal simulation time of the aforementioned operation is smaller than the time $\epsilon^{-1}=625$ considered in the work~\cite{Trotter_steps}, where $\epsilon$ is the chemical accuracy $1.6mHa$. 

We estimate the maximal simulation time by summing over the absolute value of $x_j=(j+1 / 2) D / N$ from $j=-N$ to $N-1$, which results to $\frac{\pi D}{ \Delta}$. The proxy depth $D$ is a hyper-parameter in the optimization of the booster and is set to $10$. The spectral gap $\Delta$ is equal to $[ 1.1580, 2.7287, 4.48, 6.35 ] $ for the hydrogen chains with $n=[2,4,6,8]$, which results to the maximal simulation time $[27, 12, 7, 2]$, respectively. This suggests that the maximal simulation time is smaller than the simulation time considered in the work of~\cite{Trotter_steps} and the results presented there are applicable in this analysis.

The authors present an empirical estimation of the number of Trotter steps necessary for the linear hydrogen chains with bond distance $d=1.7$ and find that approximately $K=10$ number of Trotter steps are required. According to the work of~\cite{Trotter_Error}, increasing the bond distance decreases the Trotter error, and therefore, in our resource estimations of linear hydrogen chains with $d=3.0\AA$ the empirical calculation of $K=10$ Trotter steps is a good proxy (see Fig.~\ref{fig:Hn_ratios}, Table~\ref{table:RE_booster}).

\begin{figure}
    \centering
    \includegraphics[width=1.0\linewidth]{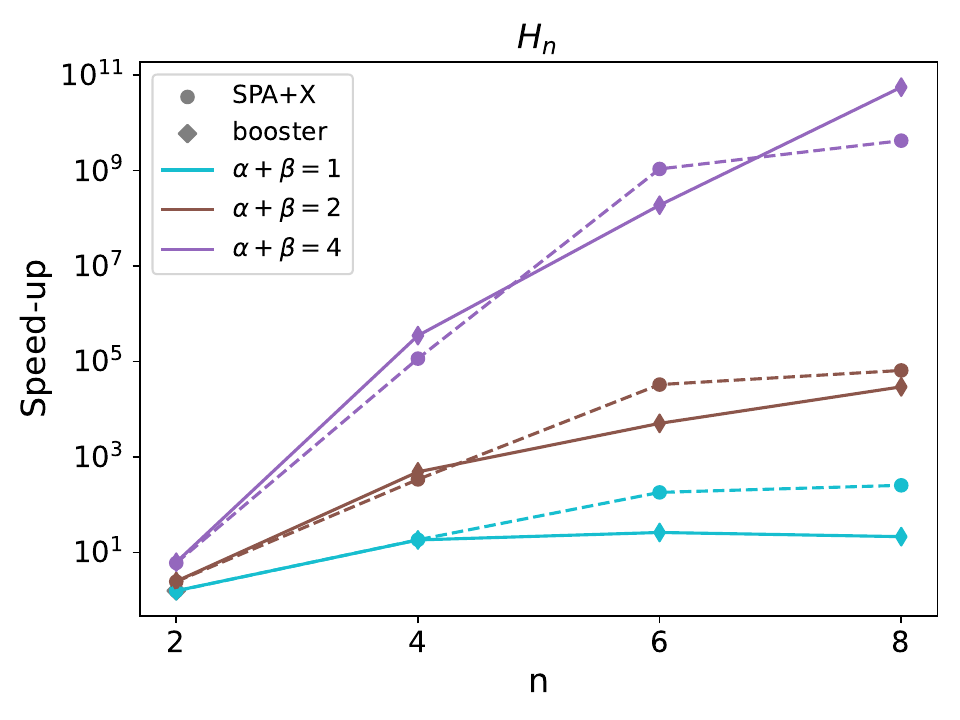}
    \caption{Here we plot the ratios $T_o/T$ for SPA and booster over HF with $K=100\sqrt{n}$ of $H_n$ system for an increasing system size $n=[2,4,6,8]$ and bond distance $d=3.0\AA$.}
    \label{fig:Hn_ratios_N100sqrt}
\end{figure}

Finally, the authors of the work~\cite{Trotter_steps} observe a $10 \sqrt{n}$ reduction in the number of Trotter steps required compared to a rigorous bound analysis given in~\cite{rigorous_bound}. In Fig.~\ref{fig:Hn_ratios_N100sqrt}, we present the runtime ratios while taking into account the increased number of Trotter steps $K=100\sqrt(n)$ for the booster algorithm instead of $K=10$ in Fig.~\ref{fig:Hn_ratios}.

\section{Overlap values}\label{App:overlaps}

In Table~\ref{table:overlaps} we presented the overlap squared values used in Fig.~\ref{fig:Hn_SPA}.

\begin{table}[ht]
    \centering
    \caption{We present overlap squared values used in Fig.~\ref{fig:Hn_SPA} for $H_n$.}
    \label{table:overlaps}
    \begin{tabular}{|c|c|c|c|c|c|c|}
        \hline
        $H_n$ & $\gamma^2_0$     &  $\gamma^2_{SPA+X}$      &  $\gamma^2_b$      \\
        \hline
        $H_2$ &  $0.63$ &   $ 0.99$              &   $1.0$                  \\
        \hline
        $H_4$ & $0.0375$ &   $0.69$ &  $1.0$       \\
        \hline
        $H_6$ & $0.0052$ & $0.94$    &   $1.0$     \\
        \hline
        $H_8$ & $0.00073$ &  $0.19$   &  $1.0$     \\
        \hline
    \end{tabular}
\end{table}

\section{Failure tolerance}~\label{App:failure_tolerance}

Usually $\delta$ is chosen through the maximal probability with which ones allow the algorithm to fail. Therefore, we define the failure tolerance $\delta_C$ as
\begin{equation}
\delta_C = 1- \left( 1-\delta \right)^R \approx R \delta \Rightarrow \delta \approx \delta_C \times \left( R \right)^{-1},
\end{equation}
where $R$ is the number of Pauli Rotations and $\delta$ is the necessary precision for operating the $R$ number of gates.

\section{Computational details}~\label{app:comp_details}

We compiled the SPA-variant circuits of Fig.~\ref{fig:Hn_SPA} to Pauli Rotations $R_{SPA}$ and CNOT gates by using the circuit compilation incorporated in \textsc{tequila}~\cite{tequila}. In Table~\ref{table:comp_details}, we present the number of Pauli Rotations $R_{SPA}$ and the respective values of $\delta_{SPA}$ for the T-gate counts of the SPA-variant circuits also presented in Table~\ref{table:RE}.

\begin{table}[ht]
    \centering
    \caption{We present the Pauli rotations ($R_{SPA}$, $R_{SPA+GS}$), the necessary precision for operating the aforementioned number of gates ($\delta_{SPA}, \delta_{SPA+GS}$)
    of the SPA-variants circuits for $H_n$.}
    \label{table:comp_details}
    \begin{tabular}{|c|c|c|c|c|}
        \hline
        $H_n$ & $R_{SPA}$     &  $\delta_{SPA}$      &    $R_{SPA+GS}$     &  $\delta_{SPA+GS}$\\
        \hline
        $H_2$ &  $1$ &   $ 10^{-3}$              &   $21$                      & $5\times10^{-5}$       \\
        \hline
        $H_4$ & $2$ &   $5 \times 10^{-4}$ &  $2.7 \times 10^3$        & $8.3 \times 10^{-6}$     \\
        \hline
        $H_6$ & $3$ & $3 \times 10^{-4}$    &   $4.2 \times 10^3$     & $3.3 \times 10^{-6}$     \\
        \hline
        $H_8$ & $560$ &  $2.5 \times 10^{-4}$   &  $2.2 \times 10^4$     & $1.8 \times 10^{-6}$  \\
        \hline
    \end{tabular}
\end{table}

For estimating the number of T-gate counts of the booster algorithm presented in Table~\ref{table:RE_booster}, we used a Trotter decomposition with $K=1$ number of steps for the Hamiltonians of $H_n$ by using the time evolution function in \textsc{orquestra}~\cite{OrquestraSDK}.
Then, we exported the circuit with Qiskit~\cite{Qiskit} and computed the number of Pauli Rotations $R_b$ with their respective values of $\delta_b$ (presented in the Table~\ref{table:comp_details} above).

\begin{table}[ht]
    \centering
    \caption{We present the Pauli rotations ($R_{b}$), the necessary precision for operating the aforementioned number of gates ($\delta_b$) and T-gate counts of the booster algorithm ($T_{B}$) for $H_n$.}
    \label{table:comp_details_booster}
    \begin{tabular}{|c|c|c|c|c|}
        \hline
        $H_n$ &  $R_b$               &  $\delta_b$               &  $T_{B}$\\
        \hline
        $H_2$ &    $9.3 \times 10^2$            & $3.3 \times 10^{-5}$       &  $1.9 \times 10^5$  \\
        \hline
        $H_4$ &  $1.3 \times 10^7$              & $7.6 \times 10^{-7}$      &  $1.3 \times 10^7$  \\
        \hline
        $H_6$ &  $7.4 \times 10^3$      & $1.4 \times 10^{-7}$                 &  $5.2 \times 10^7$ \\
        \hline
        $H_8$ &  $2.5\times 10^5 $     & $4.0 \times 10^{-8}$                 &  $1.6 \times 10^8$\\
        \hline
    \end{tabular}
\end{table}

Classical FCI and HF energies are computed with \textsc{pyscf}~\cite{pyscf}. MRA-PNOs are computed with \textsc{madness}~\cite{madness} using the implementation described in~\cite{Kottmann_2021, kottmann2020direct} on top of the framework described in~\cite{harrison2014basic, bischoff2014regularizing}. Exact diagonalization of Hamiltonians was performed with sparse solvers implemented in ~\textsc{scipy}~\cite{scipy}.
The quantum simulation backend was \textsc{qulacs}~\cite{qulacs}.

\end{document}